\documentclass[aps,pre,twocolumn]{revtex4}

\usepackage{graphicx}
\usepackage{amsmath}

\begin{document}

\title{Resonances in dissipative optomechanics with nanoparticles:\\ Sorting, speed rectification and transverse cooling}

\author{S.~J.~M.~Habraken$^{1,2,3}$}
\email{steven.habraken@uibk.ac.at}
\author{W. Lechner$^{1,2}$}
\author{P. Zoller$^{1,2}$}

\affiliation{$^1$Institute for Quantum Optics and
Quantum Information, Austrian Academy of Sciences, 6020 Innsbruck, Austria}

\affiliation{$^2$Institute for Theoretical Physics,  University of Innsbruck,
6020 Innsbruck, Austria}

\affiliation{$^3$Institute for Theoretical Physics, University of Erlangen-N\"urnberg, Staudtstra\ss e 7, 91058 Erlangen, Germany}

\date{\today}

\begin{abstract}
The interaction between dielectric particles and a laser-driven optical cavity gives rise to both conservative and dissipative dynamics, which can be used to levitate, trap and cool nanoparticles. We analytically and numerically study a two-mode setup in which the optical potentials along the cavity axis cancel, so that the resulting dynamics is almost purely dissipative. For appropriate detunings of the laser-drives, this dissipative optomechanical dynamics can be used to sort particles according to their size, to rectify their velocities and to enhance transverse cooling.
\end{abstract}

\maketitle

\noindent

\section{Introduction}

In recent years, optically levitated dielectric nanospheres \cite{CIRAC,CHANGZOLLER} have attracted considerable interest as a novel and versatile alternative to more conventional optomechanical setups \cite{KIPPENBERG, REVIEWOPTOMECHANICS, SHORTWALK, MARQUARDTREVIEW}. A dielectric object inside a cavity with fixed mirrors gives rise to dispersive coupling between the center-of-mass motion of the object and the cavity mode, similar to the interaction between a cavity mode and a moving mirror \cite{HARRIS}. Optically \cite{CIRAC, CHANGZOLLER} or magnetically \cite{SUPERCONDUCTING} levitated particles have the additional advantage that they are naturally well isolated from their environment. Combined with laser trapping and cooling, this may allow for cooling close to the quantum-mechanical ground-state \cite{GROUNDSTATE}, which offers the prospect to reach an entirely new regime in the ongoing pursuit to create quantum superpositions of particles with large masses \cite{LARGEQUANTUM,ARNDT} and to study and probe unexplored regimes of quantum mechanics and possible deviations from it \cite{COLLAPSE,PLANCKSCALE}. 

%---------------------------------------------------------------------
\begin{figure}[!t]
\centerline{\includegraphics[width=8cm]{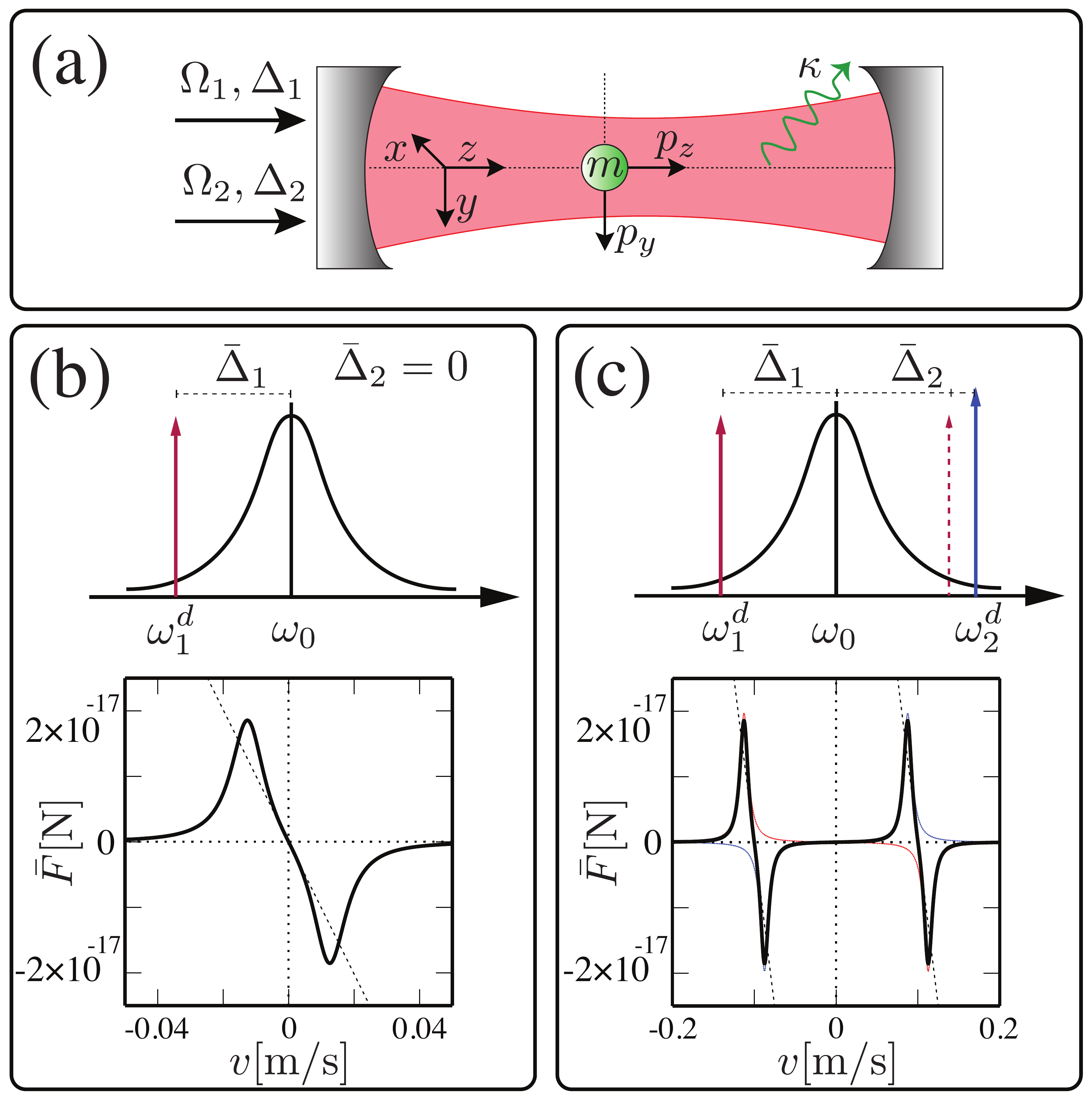}}
\caption{(a) A sub-wavelength nanoparticle of mass $m$ is confined to the axis ($z$) of an optical cavity but can move freely along it. It is optomechanically coupled to two laser-driven cavity modes. The strengths are denoted by $\Omega_{1,2}$ and the detuning from the cavity frequency is $\Delta_{1,2}$. The decay of the cavity is $\kappa$. (b) Effective damping force for one drive detuned to the red (see main text for the parameters used). For small velocities, the effective force exerted on the nanoparticle depends linearly on the velocity (dotted line) while resonances appear for larger velocities. (c) By detuning one drive to the red and the other to the blue with an additional relative detuning $\Delta_1 = -\Delta_2 + \delta$, the competing damping (dotted red) and heating (dotted blue) give rise to an effective dissipative force (black) with finite stationary velocities. Remarkably, since the force has a negative slope around these points, the stationary points are stable and all particles end up with the same final speed (determined by the detunings).}
\label{fig:illustration}
\end{figure}
%---------------------------------------------------------------------

Laser cooling in optomechanical systems has been studied extensively in the resolved-sideband regime \cite{COOLINGWILSONRAE, COOLINGMARQUARDT,RITSCHREV,RITSCHATOM}, which is analogous to the laser cooling of trapped ions \cite{IONREVIEW}, and may be applied to cool molecules \cite{RITSCHMOLECULES}. In this regime, the nanoparticle is harmonically confined to half-a-wavelength of an optical trapping mode and laser-cooled through a second cavity mode \cite{CHANGZOLLER,TWOMODES}. Recently, the interaction between a nanoparticle and a cavity mode has been considered in a more general form \cite{CKLAW, FREECOOLING}. However, laser cooling, velocity-dependent optical forces and dissipative dynamics in the regime in which the particle is not trapped but rather moves almost freely along the cavity axis, have not been explored yet. 

In this paper, we consider a setup with a nanosphere inside an optical cavity with two laser-driven modes, see Fig. \ref{fig:illustration}(a). The optical modes create a strong potential minimum along the cavity axis $z$. We consider two cases: one in which the particles are injected in the cavity so that they will be trapped and move only along the cavity axis and one in which the particles are shot through the cavity along the $x$ or $y$ axes and the velocity component along the cavity axis $z$ is optically cooled during the passage time. In both cases, the spatial directions remain uncoupled and the dynamics is effectively one dimensional. We choose the modes and drive strengths such that the conservative optical forces along the cavity axis almost completely cancel throughout the focal region of the cavity, and the resulting dynamics of the nanoparticle is dominated by dissipation.

Laser cooling in this setup is reminiscent of Doppler cooling of moving atoms \cite{MINOGIN}, but it is also fundamentally different. An atom has discrete internal levels, which makes the effective force exerted by a laser beam strongly depend on the optical frequency and, through the Doppler effect, on its speed. Dielectric nanoparticles, on the other hand, do not have any resolved internal structure and, in our setup, dissipation is purely due the finite cavity response time, which leads to a retarded, and, therefore, velocity-dependent, radiation-pressure feedback force. As one may expect, red-detuned laser drives (with respect to the bare cavity resonances) lead to damping (cooling), while blue detunings give rise to heating. 

In a two mode setup this naturally leads to two different regimes: (i) in which the particle motion is either damped or amplified by the optical modes and (ii) in which the motion is damped by one mode and amplified by the other, see Fig. \ref{fig:illustration}(b)-(c). 

In case (i) with red detunings and in the limit of small velocities, the effective damping force depends linearly on the particle momentum, see Fig. \ref{fig:illustration}(b), and gives rise to tunable spatially non-uniform Langevin dynamics \cite{DUMBBELL}. For larger momenta, we find that the effective dissipative force has a much richer structure with resonances that give rise to strongly enhanced damping, at least for specific momenta. The range of velocities for which such laser cooling works is significantly enhanced by the cancellation of the optical potential. We show that this can be relevant for quantum interference experiments in which nanoparticles are shot through a cavity and the transverse motion is optically cooled during the passage time \cite{ARNDT}. In this setup, transverse cooling is enhanced by almost two orders of magnitude.

In case (ii), the dynamics is governed by the interplay between heating and cooling, see Fig. \ref{fig:illustration}(c). If the two detunings are equal but opposite and the drive strengths are equal, the effects cancel and the motion is frictionless. However, if the magnitude of the detunings is (slightly) different, the non-linear nature of the effective forces lead to finite stationary velocities. For properly chosen detunings, these points correspond to stable stationary solutions. Initially slower particles are accelerated, while fast ones are decelerated to the same final speed. The effective dissipative force depends on the particle mass, but the stationary points in case (ii) do not. Using (ii) to initialize the particles at the same speed and, subsequently, (i) to achieve mass-dependent damping, dielectric particles can be weighted and sorted according to their mass. In this setup, a mass difference on the order of $\Delta m \simeq 10^{-19}\ {\rm kg}$ translates into a position difference on the order of $\Delta z\simeq 10^{-3}\ {\rm m}$.

This paper is organized as follows. In the next section, we introduce the setup in full detail and derive the equations of motion. Numerical results for the full model are discussed in Sec. III, while an approximate analytical treatment is presented in Sec. IV. The analytical results are in excellent agreement with the numerical simulations and provide additional insights in the nature of the dissipative optical force and its possible application to velocity-rectification, sorting and laser cooling. In Sec. V, we discuss these application in more detail and a brief summary and some concluding remarks are contained in Sec. IV.

\section{Model and Hamiltonian}
We consider the setup shown in Fig. \ref{fig:illustration}(a), in which a sub-wavelength dielectric particle interacts with two laser-driven cavity modes. For moderately strong drives, the particle is confined to the cavity axis ($z$) and we assume that the variation of the mode functions in the transverse directions ($x$ and $y$) is negligible compared to that along the $z$ axis. In this case the dynamics is effectively one dimensional and, in a frame rotating with the drive frequencies, the classical Hamiltonian is given by
\begin{equation}
\label{fullham}
H=\frac{p^2}{2m}-\sum_{j=1,2}\bigg(\Delta_{j}(z)|a_{j}|^2-\Omega_{j}(a_{j}+a_{j}^{\ast})\bigg)\;,
\end{equation}
where $j$ runs over the cavity modes, $a_{1,2}$ are classical normal variables characterizing their dynamics, $\Omega_{1,2}$ are the drive strengths and $\Delta_{1,2}(z)=\omega_{1,2}^{(d)}-\omega_{1,2}(z)$, with $\omega_{1,2}^{(d)}$ the drive frequencies, are the detunings from the mode frequencies. The optomechanical interaction between the particle and the cavity modes, derives from a dependence of the mode frequencies on the particle position $z$ along the cavity axis. We focus on the case of two modes separated by the a free spectral range of the cavity $\Delta\omega=|\omega_{1}^{(0)}-\omega_{2}^{(0)}|=c\pi/L$, where $L$ is the cavity length and $\omega_{1,2}^{(0)}$ are the frequencies in the absence of the particle. We assume that $\Delta\omega$ is much smaller than the absolute frequencies $\omega_{1,2}^{(0)}$, but much larger than any other time scale involved. In this case the mode functions near the focal region of the cavity can be approximated by $\mathbf{F}_{1}(z)\simeq\epsilon_{1}\sqrt{2/C}\sin(kz)$ and $\mathbf{F}_{2}(z)\simeq\epsilon_{2}\sqrt{2/C}\cos(kz)$ respectively. Here $C=(\pi/4)Lw^{2}$ is the mode volume, with $w$ the beam width, and $\epsilon_{1,2}$ are polarization vectors in the $xy$ plane. The resonance frequencies are modulated by the intensity at the particle position (see appendix) so that $\omega_{1}(z)=\omega_{1}^{(0)}-g_{1}\sin^{2}(kz)$ and $\omega_{1}(z)=\omega_{2}^{(0)}-g_{2}\cos^{2}(kz)$, where $g_{j}=\omega_{j}^{(0)}\alpha_{p}V/C$, with $V$ the volume of the particle and $\alpha_{p}$ its electric polarizability, the optomechanical coupling strengths. Since $\Delta\omega<<\omega_{1,2}^{(0)}$ we assume that $g_{1,2}=g$. It follows that the position-dependent detunings are given by $\Delta_{1}(z)=\bar{\Delta}_{1}-(g/2)\cos(2kz)$ and $\Delta_{2}(z)=\bar{\Delta}_{2}+(g/2)\cos(2kz)$, where $\bar{\Delta}_{1,2}=\omega_{1,2}^{(d)}-\omega_{1,2}^{(0)}\mp g/2$ are the average detunings.

The decay of the optical modes is included at the level of the equation of motion $\dot{a}_{j}=(i\Delta_{j}(z)-\kappa)a_{j}-i\Omega_{j}$. The drive term can be eliminated by moving to a shifted frame $a_{j}\rightarrow\alpha_{j}+a'_{j}$, where $\alpha_{j}=\Omega_{j}/(\bar{\Delta}_{j}+i\kappa)$ is the average amplitude in the cavity modes. The equations of motion for the residual optical amplitudes $a'$ are given by
\begin{equation}
\label{opteqmo}
\dot{a}_{j}=(i\bar{\Delta}_{j}-\kappa)a_{j}+i(\Delta_{j}(z)-\bar{\Delta}_{j})(\alpha_{j}+a_{j})\;,
\end{equation}
where we have dropped the primes. The average mode amplitudes also enter the Hamilton equations for the particle motion, which are found as
\begin{equation}
\label{parteqmo}
\dot{z}=\frac{p}{m}\qquad\mathrm{and}\qquad\dot{p}=\sum_{j}|\alpha_{j}+a_{j}|^2 \frac{\partial\Delta_{j}(z)}{\partial z}\;.
\end{equation}
The $|\alpha_{j}|^2$ terms do not depend on the optical amplitudes $a_{j}$ and describe and optical potential $V(z)=g(|\alpha_{1}|^2-|\alpha_{2}|^2)\sin(2kz)$. For $g>0$, the particle experiences a force toward the high intensity regions of the optical field. This is the basic principle underlying optical tweezers \cite{GRIER}. The optical potential $V(z)$ obviously vanishes for equal intensities $|\alpha_{1}|^2$ and $|\alpha_{2}|^2$. Since $|\alpha_{j}|^2=|\Omega_{j}|^2/(\bar{\Delta}_{j}^{2}+\kappa^2)$, cancellation of the potentials requires that the drive strengths $\Omega_{1,2}$ depend on the detunings $\bar{\Delta}_{1,2}$. Such cancellation is only exact for the approximate mode functions $\mathbf{F}_{1,2}(z)$ introduced above. This is always an approximation and, in a typical setup, the strength of the effective optical potential depends on $z$ according to $\cos^{2}(\pi z/L)$, which vanishes near $z\simeq L/2$. The potentials only cancel along the cavity axis. In the $x$ and $y$ directions, the optical fields create a potential minimum of $-g(|\alpha_{1}|^2+|\alpha_{2}|^2)$ along the cavity axis. The other terms in Eq. (\ref{parteqmo}) involve $a_{j}$ and couple the particle motion to the dynamics of the optical modes. The cavity decay $\kappa$ sets a finite time scale for the cavity field to adopt to changes in the particle position, so that the radiation-pressure force on the particle is retarded and not only depends on the particle position, but also on the momentum along the cavity axis. Depending on the sign of the detuning, such forces give rise to damping ($\bar{\Delta}<0$) or heating ($\bar{\Delta}>0$). We focus on the case in which $|\alpha_{1}|=|\alpha_{2}|$ so that, at least to leading order, the optical potentials cancel, and the dynamics is dominated by dissipation.

\section{Numerical results}
\label{sec:numerics}
Before we solve the equations of motion (\ref{opteqmo}) and (\ref{parteqmo}) analytically by making further assumptions, we first study the full dynamics numerically. We assume that a single particle is carefully injected in the cavity so that it is trapped and, initially, moves along the cavity axis at a constant speed $v(0)$. We numerically solve (\ref{opteqmo}) and (\ref{parteqmo}) for the $z(t)$ and $v(t)=p(t)/m$ for different particle masses. For concreteness, we consider Silica ($\rho=2200\ {\rm kg/m^{3}}$ and $\epsilon_{r}=2.34$) nanospheres with radii around $R=50\ {\rm nm}$, which have masses around $m=1.15\times 10^{-18}\ {\rm kg}$. The optomechanical coupling strength $g$ is proportional the particle volume (or mass) and, for a cavity length of $L=0.12\ {\rm m}$ and a mode width of $w=100\ {\rm \mu m}$, we find optomechanical coupling strengths around $g\simeq 10^{3}\ {\rm Hz}$. Furthermore, we choose a wavelength $\lambda=10^{-6}\ {\rm m}$ for both modes and a cavity finesse of $\mathcal{F}=50\times 10^{3}$, which corresponds to a field decay rate of $\kappa=\pi c/(2L\mathcal{F})\simeq 8\times 10^{4}\ {\rm Hz}$. The intra-cavity power is taken to be $P_{\rm cav}\simeq 50\ {\rm W}$ per mode so that $|\alpha_{1,2}|\simeq 3.2\times 10^{-12}\ \sqrt{Js}$. For a $50\ {\rm nm}$ particle, the optical confinement to the cavity axis $-g(|\alpha_{1}|^{2}+|\alpha_{2}|^{2})$ corresponds to a velocity of $0.19\ {\rm m/s}$.

%---------------------------------------------------------------------
\begin{figure}[t]
\centerline{\includegraphics[width=8cm]{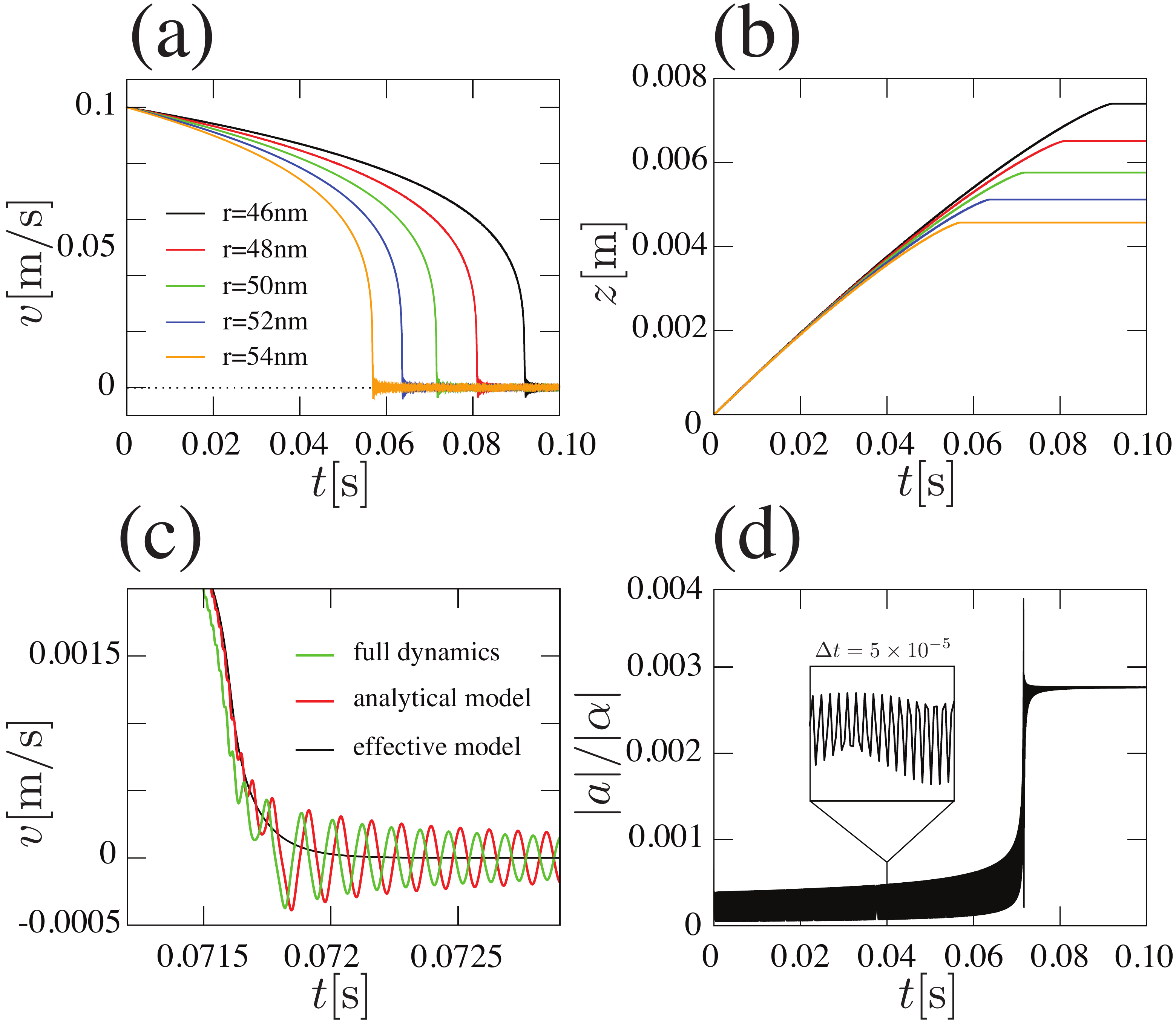}}
\caption{(a) Velocity as a function of time for various nanosphere sizes ranging from $r=46\ {\rm nm}$ (black) to $r=54\ {\rm nm}$ (orange).  (b) Position as a function of time for the same nanospheres. All particles are eventually trapped, but the distance traveled strongly depends on the size. (c) Zoom in to the point where a particle is trapped in the small reminiscent optical potential. The result from the effective force Eq. (\ref{avforce}) (black), the analytic result from Eq. (\ref{fullforce}) (green) and the full numerical data (green) are in excellent agreement up to the point where the particle is trapped. (d) The amplitude of the cooling mode $|a_{1}|/|\alpha|$ as function of time for $R=46\ {\rm nm}$. The regime of strongly enhanced cooling in (a) corresponds to a sharp resonance in the optical response.}
\label{fig:sorting}
\end{figure}
%---------------------------------------------------------------------

For these parameters and with the first cavity mode detuned to the red $\bar{\Delta}_{1}=-2\kappa\simeq -1.6\times 10^{5}\ {\rm Hz}$ and the other driven resonantly $\bar{\Delta}_{2}=0$, we numerically simulate the cooling and trapping of different nanospheres, all with the same initial speed $\dot{z}(0)=0.10\ {\rm m/s}$. The time dependent position and the speed of the particles are respectively shown in Fig. \ref{fig:sorting}(a) and (b). All particles are strongly cooled, and, eventually, trapped in some reminiscent small harmonic potential, as can be seen in Fig. \ref{fig:sorting}(c). This takes place within a tenth of a second and within a distance of a few millimeter. Both the time it takes to be trapped and the distance traveled strongly depend on the particle mass, so optomechanical cooling in this form allows to sort particles according to their mass. The particle speeds shows a sharp jumps around $v\simeq\bar{\Delta}_{j}/(2k)=1.25\times 10^{-2}\ {\rm m/s}$ for which we observe strongly enhanced cooling. These jumps are obviously due to resonances, as can be seen from the amplitude in the cooling mode $|a_{1}|/|\alpha|$, as plotted in Fig. \ref{fig:sorting}(d). The optical amplitude slightly shifts once the particle is trapped. This is due to the fact that the particle is trapped at a potential minimum for which the optical resonance frequency differs slightly  from that for a moving particle.

In order to use the mass-dependent cooling and trapping to weight and sort particle according to their mass, we also need to initialize all particles to the same speed. This can be achieved by combining optomechanical damping and heating in our two-mode set up. In Fig. \ref{fig:rectification}, we show full numerical results for the parameters given above, $\bar{\Delta}_{1}=-(4\pi\times 10^{5}+2\kappa) =-1.4\times 10^{6}\ {\rm Hz}$ and $\bar{\Delta}_{2}=4\pi\times 10^{5}-2\kappa=1.1\times 10^{6}\ {\rm Hz}$ and various initial speeds $\dot{z}(0)$ ranging from 0.02 m/s to 0.16 m/s. The results show that all particles reach the same velocity of $0.1\ {\rm m/s}$ on the time scale of less than a hundredth of a second, which, again, corresponds to a traveled distance of few millimeters.

\section{Analytical results}
In order to gain some insight in these features, we analytically study the interaction between a single nanoparticle and the cavity modes. This analysis is general and, below, we shall apply it to various parameter regimes and applications. 

\subsection{Effective optical force}
In order to derive an expression for the effective optical force exerted on the particle, we make two simplifications: (a) if the second term in Eq. ({\ref{opteqmo}}), which describes excitation of the optical mode through the optomechanical coupling, is much smaller than the first, the optical modes can be adiabatically eliminated and (b) if $a_{j}<<\alpha_{j}$, the optomechanical interaction can be linearized, i.e. $(\Delta_{j}(z)-\bar{\Delta}_{j})(\alpha_{j}+a_{j})$ in (\ref{opteqmo}) can be approximated by $(\Delta_{j}(z)-\bar{\Delta}_{j})\alpha_{j}$ and $|\alpha_{j}+a_{j}|$ in (\ref{parteqmo}) by $|\alpha_{j}|^2+\alpha^{\ast}_{j}a_{j}+\alpha_{j}a_{j}^{\ast}$.

As a first step to eliminating optical modes, we specify initial conditions at $t\rightarrow-\infty$, choose $a_{j}(-\infty)=0$ and formally solve the linearized form of (\ref{opteqmo})
\begin{equation}
\label{adel}
a_{j}(t)=i\alpha_{j}\int_{-\infty}^{t}dt'\;e^{(i\bar{\Delta}_{j}-\kappa)(t-t')}\bigg(\Delta_{j}(z(t'))-\bar{\Delta}_{j}\bigg)\;.
\end{equation}
Substitution of this result in (\ref{parteqmo}) yields a closed description in terms of $z$ and $p$, which no longer involves the optical modes but, since the integrand in (\ref{adel}) depends on $z(t')$, these equations involve integration of $z(t')$ and are non-local in time. However, since $(\Delta_{j}(z(t'))-\bar{\Delta}_{j})\sim g$, it follows from (\ref{adel}) that $a_{j}\sim \alpha_{j} g/(i\bar{\Delta}_{j}-\kappa)$, which in (\ref{parteqmo}) is multiplied by $\partial\Delta_{j}/\partial z\sim gk$ so that, to second order in $g$, if suffices to solve (\ref{parteqmo}) to zeroth order and substitute the result in (\ref{adel}). With $|\alpha_{1}|=|\alpha_{2}|$ the equations of motion (\ref{parteqmo}) to zeroth order describe a free particle, so that $z(t')\simeq z(t)+(p(t)/m)(t'-t')$ and $\Delta_{j}(z(t'))-\bar{\Delta}_{j}\simeq \mp g\sin^{2}\left(k\left(z(t)+\frac{p(t)}{m}(t'-t)\right)\right)$, where the minus sign refers to the first mode. At this level of approximation, the modes are not coupled and the optical forces exerted on the particle remain independent. Coupling of the modes only appears in $\mathcal{O}(g^4)$. After substitution of $\Delta_{j}(z(t'))$ in (\ref{adel}), the integral can be evaluated

\begin{widetext}
\begin{eqnarray}
a_{j}=\mp \frac{\alpha_{j} gm}{4} \Bigg(\frac{e^{2ikz}m}{2kp-m(\bar{\Delta}_{j}+i\kappa)}-\frac{e^{-2ikz}m}{2kp+m(\bar{\Delta}_{j}+i\kappa)}\Bigg)
\end{eqnarray}
and, for $|\alpha_{1}|=|\alpha_{2}|=|\alpha|$, the effective forces are found from (\ref{parteqmo})
\begin{eqnarray}
\label{fullforce}
F_{j}=\frac{g^2|\alpha|^2\bar{\Delta}_{j}km\sin(2kz)}{2}\Bigg(
\frac{m\kappa\sin(2kz)-(2kp-m\bar{\Delta}_{j})\cos(2kz)}{(2kp-m\bar{\Delta}_{j})^2+m^2\kappa^2}
\qquad\qquad\qquad\quad\qquad\qquad\qquad\qquad\nonumber\\
-\frac{m\kappa\sin(2kz)-(2kp+m\bar{\Delta}_{j})\cos(2kz)}{(2kp+m\bar{\Delta}_{j})^2+m^2\kappa^2}
\Bigg)\;.
\end{eqnarray}
If the particle is not trapped and the momentum change per wavelength traveled is sufficiently small, the effective damping force can be obtained by averaging over a wavelength
\begin{equation}
\label{avforce}
\bar{F}_{j}(p)=\frac{k}{2\pi}\int_{0}^{2\pi/k}dz\; F(z,p)=
\frac{g^2|\alpha|^2 km^2\kappa}{4}\left(\frac{1}{(2kp-m\bar{\Delta}_{j})^{2}+m^2\kappa^2}-\frac{1}{(2kp+m\bar{\Delta}_{j})^{2}+m^2\kappa^2}\right)\;.\end{equation}
\end{widetext}
If, on the other hand, the particle momentum approaches $0$, the effective force (\ref{fullforce}) approaches $\tilde{F}_{j}(z)=g^{2}|\alpha|^2 \bar{\Delta}_{j}k\sin(4kz)/(2(\bar{\Delta}_{j}^{2}+\kappa^2)$, which corresponds to an effective potential
\begin{equation}
\label{effpot}
\tilde{V}_{j}(z)=\frac{g^{2}|\alpha|^2 \bar{\Delta}_{j}\cos(4kz)}{8(\bar{\Delta}^{2}+\kappa^2)}\;.
\end{equation}
Although the bare optical potentials cancel if $|\alpha_{1}|=|\alpha_{2}|$, in regime (i), cooled particles are eventually trapped due these small additional potentials. Once trapped, the particles will perform damped oscillations around their final positions. Similarly, in the combined damping and heating setup (ii), the velocity will perform small velocity oscillations, even after it has reached its stationary value.

%---------------------------------------------------------------------
\begin{figure}[!t]
\centerline{\includegraphics[width=8cm]{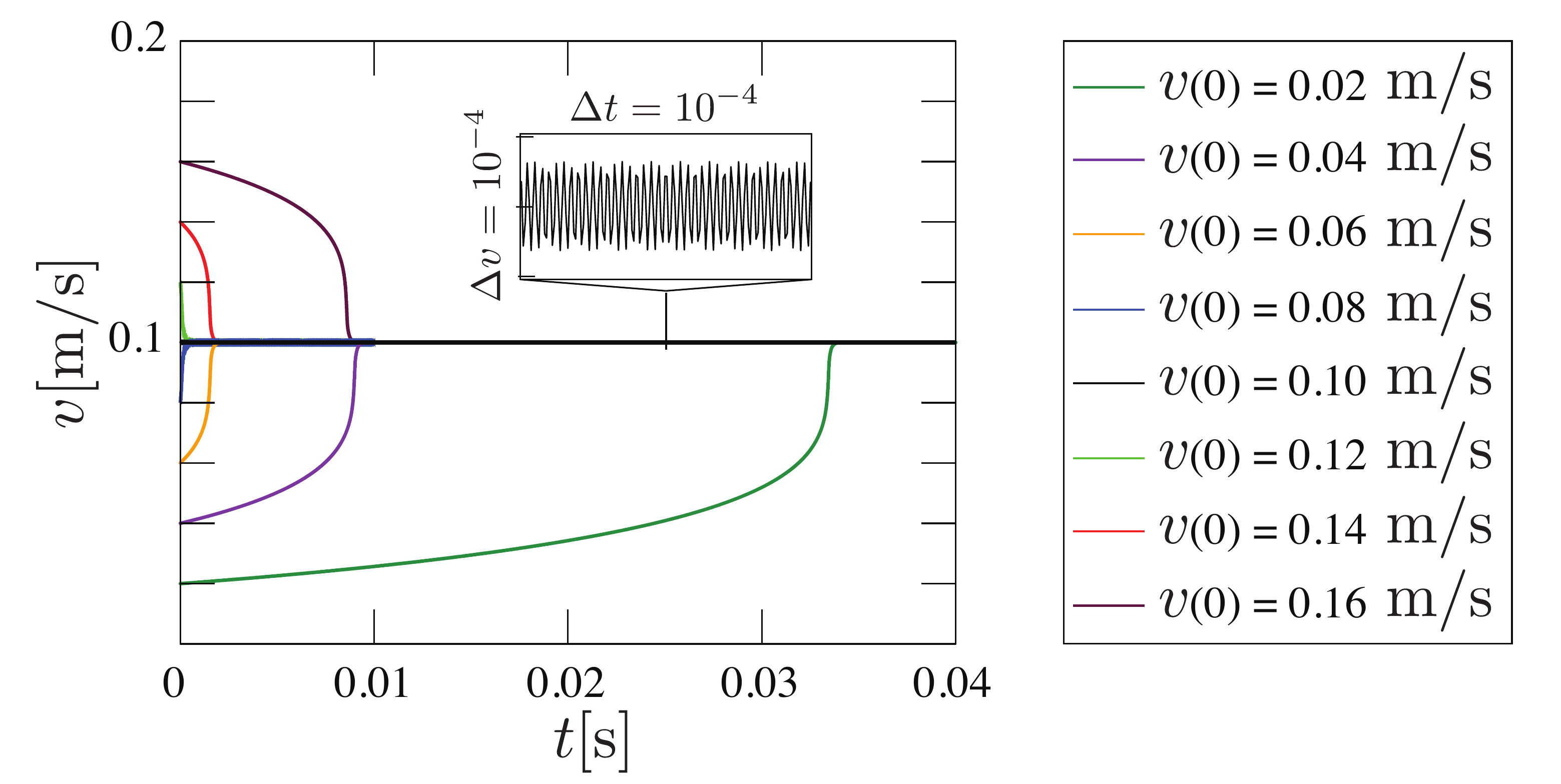}}
\caption{Velocity as a function of time for particles of size $r=50\ {\rm nm}$ for various initial velocities ranging from $v_0=0.02\ {}\rm m/s$ to $v_0=0.16\ {\rm m/s}$ in a two-mode setup with one damping and one amplifying mode. Particles with velocities below $v=0.1\ {\rm m/s}$ are accelerated while particles with larger velocities are decelerated to reach the same stationary velocity $v=0.1\ {\rm cm/s}$. }
\label{fig:rectification}
\end{figure}
%---------------------------------------------------------------------

The optical forces (\ref{avforce}) and (\ref{fullforce}) have sharp resonances around $p_{\rm res}=\pm \bar{\Delta}_{j}m/(2k)$. For these values of the momentum, the modulation of the intensity of the optical field, as perceived by the moving nanoparticle, $p/(2mk)$ is resonant with the optical detuning $\bar{\Delta}_{j}$. Such resonances give rise to sharp jumps in the velocity of the particle and correspond to regimes of strongly enhanced damping (or heating). This is illustrated in Fig. {\ref{fig:sorting}}(a)-(b) and in Fig. {\ref{fig:rectification}}. The width of the resonances is set by the optical damping rate $\kappa$ according to $\Delta p=\kappa m/(2k)$ and it follows that for $\bar{\Delta}>\kappa$, the sidebands are resolved. For $p>0$ ($p<0$) and a red detuning $\Bar{\Delta}_{j}<0$, the second (first) terms in  (\ref{fullforce}) and (\ref{avforce}) are dominant. These terms give rise to damping. For blue detunings $\bar{\Delta}_{j}<0$, it is the other way around, and the optical force gives rise to heating. Higher-order resonances appear at integer multiples of $\bar{\Delta}m/(2k)$ in higher-orders of $g^{2}$. However, for typical experimental parameters, these are highly suppressed.

Both the average (\ref{avforce}) and the effective (\ref{fullforce}) forces can easily be integrated numerically. For the parameters discussed above, the results virtually coincide with those of our full numerical simulations. This confirms that the approximations made are well justified for typical experimental parameters. As illustrated in Fig. \ref{fig:sorting}(c), somewhat larger quantitative differences only appear once the particle is trapped by the small reminiscent potentials (\ref{effpot}). At this point, the assumption that the zeroth order motion is that of a free particle is no longer a very accurate one and standard-second order optomechanical laser cooling theory \cite{COOLINGWILSONRAE, COOLINGMARQUARDT}, which assumes harmonic oscillations to zeroth order, would give a more accurate effective description. In the regime where we are interested in, all features are accurately described by the effective force (\ref{fullforce}). The effective damping force (\ref{avforce}) almost perfectly describes the trend line of the curves.

\subsection{Mass-dependence}

Having derived the effective optical force, we study how the optomechanical damping in regime (i) depends on the particle mass. The velocities at which the resonances occur $v_{\rm res}=\pm \bar{\Delta}_{j}/(2k)$ are obviously independent of mass, as is also confirmed by the numerical results in Fig. \ref{fig:sorting}(a)-(b). The particle mass only appears explicitly in $\dot{z}=p/m$ and it follows that the optical forces (\ref{fullforce}) only depend on the mass through the momentum, i.e. $F_{j}(z,p) =F_{j}(z,m\dot{z})$. However, for actual nanospheres, the optomechanical coupling strength $g$ scales with the particle volume (see appendix) and, therefore, with the mass. Since $F\propto g^{2}$, it follows that also $F_{j}\propto m^{2}$ so that only $F_{j}(z,p)/m^2$ is truly independent of mass. Thus we find from $m\ddot{z}=\sum_{j} F_{j}(z,p)$ that, for given initial speed $\dot{z}(0)$, the time it takes to be trapped scales with $1/m$. Since $z=\int dt\ \dot{z}(t)$, the same is true for the distance traveled. These conclusions are in obvious agreement with the numerical results in Fig. \ref{fig:sorting}(b).

\subsection{Optomechanical velocity control}
The combined damping and heating in regime (ii) requires a little more analysis. In order to gain some insight, we start from the effective damping force (\ref{avforce}) for both modes. For the sake of the argument, we assume that $p>0$ and, for simplicity, we neglect the off-resonant terms in (\ref{avforce}). In this case, the  damping (1) and amplifying (2) forces are respectively given by
\begin{equation}
\bar{G}_{j}(p)=\mp\frac{g^2|\alpha|^2 km^2\kappa}{4(2kp\pm m\bar{\Delta}_{j})^2+m^2\kappa^2}\;.
\end{equation}
For equal but opposite detunings $\bar{\Delta}_{1}=-\bar{\Delta}_{2}$, the total force $\bar{G}_{\rm tot}=\sum_{j}\bar{G}_{j}$ obviously vanishes. As is illustrated in Fig. \ref{fig:illustration}, this is not true for (slightly) different detunings $\bar{\Delta}_{1}=-(\Delta+\delta)$ and $\bar{\Delta}_{2}=\Delta-\delta$, in which case the total force is found as
\begin{widetext}
\begin{eqnarray}
\bar{G}_{\rm tot}=\frac{g^2|\alpha|^2 k m^{2}\kappa}{4}\Bigg(\frac{1}{(2kp-m(\Delta-\delta))^2+m^2\kappa^2}-\frac{1}{(2kp-m(\Delta+\delta))^2+m^2\kappa^2}\Bigg)\;.
\end{eqnarray}
\end{widetext}
This total force has a stationary point for $p=m\Delta/(2k)=m(\bar{\Delta}_{2}-\bar{\Delta}_{1})/k$. The slope around this point is given by
\begin{equation}
\left(\frac{\partial\bar{G}_{\rm tot}}{\partial p}\right)_{p=\frac{m\Delta}{2k}}=-\frac{2g^2|\alpha|^2 k^2 \delta\kappa}{m(\delta^2+\kappa^2)^2}
\end{equation}
and is negative for $\delta>0$, in which case the stationary point is stable. Physically speaking, this means that the red-detuned damping laser is further detuned than the blue-detuned amplifying laser. Initially too fast particles are slowed down, while initially slow ones are accelerated to reach the same stationary velocity $\dot{z}=p/m=\Delta/(2k)$, independent of their mass. Assuming that $p<0$, similar arguments lead to the conclusion that also $p=-m\Delta/(2k)$ is a stable stationary point of $\delta>0$.

From a calculation similar to the one that led to Eq. (\ref{effpot}), we find a small oscillatory potential for particles with velocities around the stationary  points
\begin{equation}
V_{\rm tot}\simeq -\frac{g^{2}|\alpha|^{2}\delta\cos(4kz)}{8(\delta^{2}+\kappa^2)}\;.
\end{equation}
This gives rise to small velocity fluctuations around the stationary velocities. For the parameters discussed in Sec. III, $\delta=2\kappa\simeq 1.6\times 10^{5}\ {\rm Hz}$ (as for the numerical results in Fig. \ref{fig:rectification}), we find $g^{2}|\alpha|^2\delta/(8(\delta^2+\kappa^2))\simeq 6.3\times 10^{-24}\ {\rm J}$, which is almost three orders of magnitude smaller than thermal fluctuations at room temperature. For a particle with mass $m=1.15\times 10^{-18}\ {\rm kg}$ and at a speed of $0.1\ {\rm m/s}$, it corresponds to velocity fluctuations of $\sim 5.4\times 10^{-5}\ {\rm m/s}$. This is in excellent agreement with the numerical result shown in the insert in Fig. \ref{fig:rectification}. We emphasize that such velocity fluctuations are, at least in principle, fully coherent. In this respect they are fundamentally different from thermal fluctuations.

\section{Applications}

\subsection{Weighting and sorting}
Combining both settings, we envision the following scheme to sort particles according to their mass. Nanoparticles of unknown mass are inserted one by one in the optical cavity and confined to the cavity axis. On a length scale of a few millimeter, they are either accelerated or slowed down to a fixed speed of $\sim 0.1\ {\rm m/s}$ using the combined optomechanical cooling and heating. Since the distance traveled during the velocity rectification depends both on the initial velocity and on the mass, the particles need to be detected when they pass a certain transverse plane of the cavity. Once a particle has been detected, the detuning of the heating mode is switched from the blue to the bare cavity resonance, so that the other mode cools the particle until it is trapped, again on the length scale of a few millimeter. Both steps easily fit within the focal range of a cavity of length $L=0.12\ {\rm m}$, where the cancellation of the optical potentials is (almost) complete. The distance traveled from the detection until the trapping is inversely proportional to the mass of a particle and the mass can be determined from the position where the particle is trapped. If the effects of thermal fluctuations are sufficiently reduced by lowering the background pressure, the measurement accuracy is limited by the position measurements. These are, at best, accurate up to $\lambda=10^{-6}\ {\rm m}$, which amounts to a relative error of the order of $10^{-3}$ in the particle mass. Switching times are limited by $\kappa/2=4\times 10^{4}\ {\rm Hz}$, which for velocities around $0.1\ {\rm m/s}$ correspond to position errors of, at most, a few wavelengths. The effect of velocity fluctuations due to the small optical potential (\ref{effpot}) amount to even smaller relative errors of the order of $10^{-4}$.

%---------------------------------------------------------------------
\begin{figure}[t]
\centerline{\includegraphics[width=8cm]{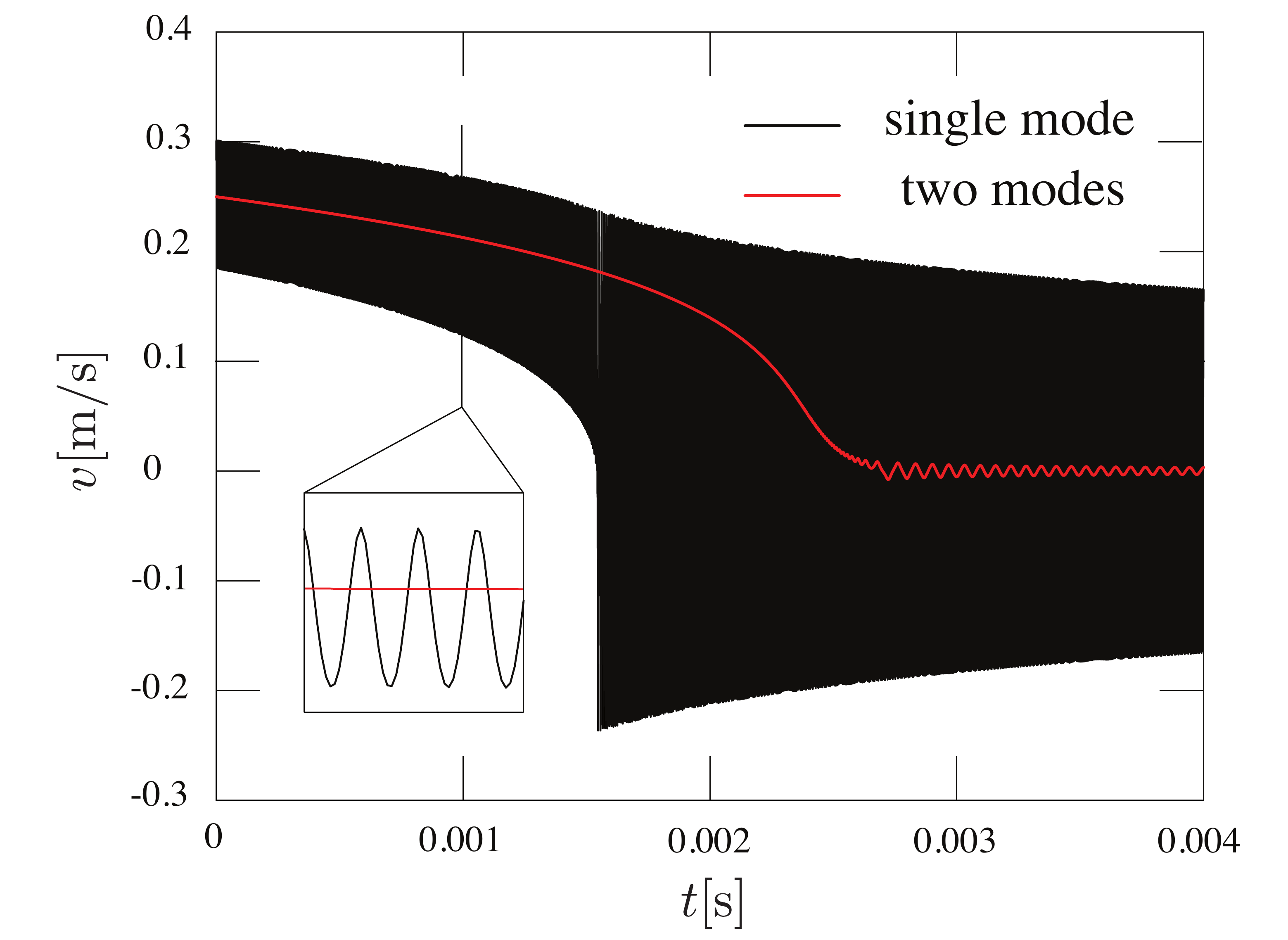}}
\caption{Laser cooling of a $50\ {\rm nm}$ nanoparticle and $\dot{z}(0)=0.25\ {\rm m/s}$ with (red) and without (black) a compensating potential. Without the compensating mode, the particle oscillates rapidly and is quickly trapped so that this type of laser cooling becomes much less efficient. Even with a compensating potential, the particle is eventually trapped, but the range of strong exponential damping is much larger. For parameters, see main text.}
\label{fig:cooling}
\end{figure}
%---------------------------------------------------------------------

For the parameters discussed here, mediated interactions between different particles are weak, and numerical simulations on two particles confirm that the small oscillatory optical potential is strong enough the keep the first particle trapped while the second one is inserted and slowed down. For beam widths much larger than the particle diameters, the cross section of collisions will be relatively small and, for just a few particles, all this could work within a single cavity. For larger particle numbers, collisions become more likely. In the spirit of Ref. \cite{LARGEQUANTUM}, once trapped, one could let particles fall down into a second storage cavity to avoid collisions.

\subsection{Transverse cooling towards the quantum regime}
The experimental parameters discussed in Sec. \ref{sec:numerics} were chosen so as to illustrate the resonances, the mass-dependent damping and velocity rectification, as shown in  Fig. \ref{fig:sorting} and \ref{fig:rectification}. They lead to rather modest overall cooling. The effective cooling rate is largest if the detunings are chosen such that the initial speed of the particle is around or just below the resonance and efficient cooling only works up to the point where the particle is trapped. In this regime of $|p|<\bar{\Delta}m/(2k)$, the effective optical damping force (\ref{avforce}) exerted by a single optical mode on a nanoparticle gives rise to an approximately constant cooling rate
\begin{equation}
\label{gammaopt}
\gamma_{\rm opt}=\frac{2k^2 g^{2}|\alpha|^2 \bar{\Delta}\kappa}{m(\bar{\Delta}^{2}+\kappa^{2})^2}\;.
\end{equation}
For the experimental parameters discussed in Sec. \ref{sec:numerics}, this equals $\gamma_{\rm opt}\simeq 8.9\times 10^{3}\ {\rm Hz}$ and gives rise the the small domains with strong exponential cooling Fig. \ref{fig:sorting}a. In principle, the cooling rate can be improved by several orders of magnitude by using smaller cavities, in which the optomechanical coupling is stronger, and higher intra-cavity powers. It may be comparable to or even exceed the cooling rate in the resolved side band regime $\sim 3\times 10^{5}\ {\rm Hz}$, as estimated in \cite{CHANGZOLLER}. However, larger $|\alpha|^2 g^2$ also enhances the residual optical potential (\ref{effpot}), so that one also has to reach a compromise between the range of velocities that is cooled and the cooling rate. Smaller cavities and larger intra-cavity powers are currently operated  to cool molecules and nanoparticles in quantum interference experiments with nanoparticles and large molecules \cite{ARNDT}. In this setup the particles are shot through the cavity, as illustrated in Fig.\ref{fig:illustration}(a) and the transverse motion is strongly cooled during the passage time.

In order to illustrate the cooling potential of our two-mode setup and the improvement compared to standard single-mode laser cooling, we consider a smaller cavity with $L=0.01\ {\rm m}$ and a larger intra-cavity power of $P=3.2\times 10^{2}\ {\rm W}$. We choose $\bar{\Delta}_{1}=-\kappa/\sqrt{3}$ so as to optimize the cooling rate ({\ref{gammaopt}}) and $\bar{\Delta}=0$ for the other, compensating mode. Leaving all other parameters unchanged, we find $\kappa\simeq 9.4\times 10^{4}\ {\rm Hz}$, which through $\bar{\Delta}_{1}=-\kappa/\sqrt{3}$ and $v_{\rm res}=\bar{\Delta}_{1}/(2k)$ corresponds to a resonant speed of $4.3\times 10^{-2}\ {\rm m/s}$. From (\ref{effpot}), we find that the particle is trapped for speeds below $8.7\times 10^{-3}\ {\rm m/s}$, which leaves a reasonably large range for exponential cooling. In the absence of the compensating mode, the particle is trapped in an effective potential of strength $g|\alpha|^2$, which corresponds to a velocity of $0.24\ {\rm m/s}$ and leaves no range for efficient cooling at all. In Fig \ref{fig:cooling}, we show full numerical results for both setups. These clearly illustrate the enhancement of laser cooling due to the compensating potential.

In principle, the range of velocities can be further increased by also compensating the effective potential (\ref{effpot}) or by adiabatically varying the detunings and drive strengths. However, eventually, the particle will always be trapped in some small optical potential and, in order to reach the motional quantum ground state, one has to apply resolved side-band cooling as well. The resonances and potential cancellation we have discussed in this paper could be used to enhance pre-cooling in the regime where the particle is not yet trapped and the strengths may be optimized so as to make the transition between the two cooling regimes as fast as possible.

\section{Summary and conclusion}
We have studied the classical dynamics of sub-wavelength dielectric nanoparticles in an optical cavity with two laser-driven modes. The modes where chosen such that the optical potentials along the cavity axis cancel and the dynamics is dominated by dissipation. In a real system, such cancellation is always an approximation, but numerical simulations have confirmed that, at least in the regime in which the particle is not trapped, a small imbalance of the optical intensities hardly affects the dynamics. We have focused on the case in which the optical intensities are strong enough to confine the particle to the cavity axis, but our results also apply to setups in which particles are shot through the cavity orthogonal to the cavity axis. In either case, the dynamics is effectively one dimensional. This setup is somewhat reminiscent of Doppler cooling of atoms but, since nanospheres do not have any resolved internal structure, it is also fundamentally different.

The availability of two resolved optical modes naturally leads to two different regimes: (i) in which the particle motion is either damped or amplified by the optical modes and (ii) in which the particle motion is damped by one mode and amplified by the other. In case (i), we find mass-dependent damping and heating, while regime (ii) gives rise to tunable stable stationary velocities. We have derived intuitive analytical expressions of the effective force on the particle, the effective dissipative force and the perturbative optical potential in which cooled particles are eventually trapped. The analytical results are in excellent agreement with full numerical simulations. This analytical treatment is accurate up to second order in the optomechanical coupling strength $g$. Since $g$ is proportional to the particle mass, it follows that, the effective optical force scales with $m^2$ so that, in case (i) and for a given initial velocity, both the time it takes for a particle to be cooled and trapped and the distance traveled are inversely proportional to the particle mass. This allows for weighting and sorting of nanoparticles according to their mass. The measurement accuracy is limited by the position measurements which are, at best, accurate up to a wavelength. For realistic parameters this allows for a relative error on the order of $10^{-3}$. In regime (ii), the combination of heating and cooling leads to stationary stable velocities, independent of the particle mass. This can be used to initialize different particles to the same speed.

The setup and approach discussed in this paper are more generally applicable. Both the mass-dependent damping (or heating) and the velocity rectification may be optimized for specific needs by adiabatically varying the detunings of the laser drives. Thus it should be possible to benefit from the resonant cooling over a larger range of velocities and to reduce the residual velocity fluctuations after rectification by another order of magnitude. Moreover, we hope that the approach discussed here will also prove fruitful in the ongoing efforts to create and manipulate coherent quantum states of dielectric nanoparticles. A full quantum treatment would require a detailed analysis of the effects of thermal decoherence and optical scattering. For the cooling, this would be very similar to the analysis in Ref. \cite{CHANGZOLLER}. Heating and, in particular, the combination of damping and heating poses additional challenges since it involves relatively large speeds and additional fluctuations. At present, it is not yet clear how these compete with quantum coherence. Obviously, both heating and combined damping and heating may have interesting applications in the quantum domain. Heating would, for instance, allow for a strong enhancement of quantum fluctuations and the combination of heating and cooling could drastically increase the velocity spread of a quantum wave packet.

\section{Acknowledgements}
It is a pleasure to thank P.~Rabl, M.~Aspelmeyer, N.~Kiesel, M.~Arndt, P.~Asenbaum and S.~Kuhn for fruitful discussions. SJMH thanks for F.~Marquardt for providing a stimulating environment during the final stages of this work. Work at Innsbruck is supported by the integrated project AQUTE and SIQS and the Austrian Science Fund through SFB F40 FOQUS.
 
\appendix

\section{Optomechanical interaction}
The interaction between a single optical cavity mode and a dielectric particle derives from a dependence of the optical resonance frequency $\omega(z)$ on the particle position $z$ along the cavity axis. With the particle inside, the displacement field in the cavity is given by $\mathbf{D}=\epsilon_{0}\mathbf{E}+\mathbf{P}$, where $\mathbf{E}$ is the optical electric field and $\mathbf{P}=\epsilon_{0}\alpha_{p}\Theta(\mathbf{r})$ is the polarization due to the particle. Here $\alpha_{p}=3(\epsilon-\epsilon_{0})/(\epsilon+2\epsilon_{0})$ is the polarizability of the particle with $\epsilon$ the permitivity \cite{JACKSON} and $\Theta(\mathrm{r})$ equals $1$ for $\mathbf{r}$ inside the particle and and $0$ elsewhere. Under the assumption that $\mu=\mu_{0}$, so that $\mathbf{H}=\mu_{0}^{-1}\mathbf{B}$, we find from the homogeneous Maxwell equations the modified wave equation
\begin{equation}
\label{waveq}
\left(\nabla^{2}-\frac{(1+\alpha_{p}\Theta(\mathbf{r}))}{c^{2}}\frac{\partial^{2}}{\partial t^2
}\right)\mathbf{E}=0
\end{equation}
with $c$ the speed of light. In the absence of the particle ($\Theta(\mathbf{r})\rightarrow 0$), this equation reduces to the standard wave equation, which, for given mirror positions and curvatures, can be solved for the complete and orthonormal set of mode functions $\mathbf{F}_{j}$ and frequencies $\omega_{j}=ck_{j}$, with $\nabla^{2}\mathbf{F}_{j}=k^{2}_{j}\mathbf{F}_{j}$ and where $j$ abbreviates spatial and polarization mode indices. If the effect of the additional term $\alpha\Theta(\mathbf{r})$  in (\ref{waveq}) is sufficiently small, it can be treated perturbatively. To first order, the mode functions are not modified. The eigenvalues $\omega_{\lambda}^2$, on the other hand, are shifted by $-\alpha_{p}\int_{C}d_{3}\mathbf{r}\;\mathbf{F}_{j}^{\dag}(\mathbf{r})\Theta(\mathbf{r})\mathbf{F}_{j}(\mathbf{r})$,
where $C$ denotes the volume of the cavity. Since the integrand differs from zero only for $\mathbf{r}$ inside the particle, the integration can be restricted to the volume of the particle. When the particle size is much smaller than the optical wavelength, the variation of the mode functions $\mathbf{F}_{j}(\mathbf{r})$ over the particle volume can be neglected and we find
\begin{equation}
\omega'_{\lambda}\simeq\omega_{\lambda}\left(1-\frac{\alpha V}{2}\left|\mathbf{F}_{j}(\mathbf{r}_{0})\right|^2\right)\;,
\end{equation}
where $\mathbf{r}_{0}$ denotes the (center-of-mass) position of the particle. For the standing-wave modes $\mathbf{F}_{1,2}(z)$ introduced in the main text, this reduces to $\omega_{1,2}(z)$.

\section{Full Hamiltonian}
In general, the vector potential corresponding to a set of cavity modes $\mathbf{F}_{j}$ is given by $\mathbf{A}=\sum_{j}\xi_{j}\mathbf{F}(\mathbf{r})$, from which the canonical momentum density is found as $\boldsymbol{\Pi}(\mathbf{r},t)=\epsilon_{0}\dot{\mathbf{A}}(\mathbf{r},t)=\epsilon_{0}\sum_{j}\dot{\xi}_{j}(t)\mathbf{F}_{j}(\mathbf{r})\equiv\sum_{j}\pi_{j}(t)\mathbf{F}_{j}(\mathbf{r})$. From the general Hamiltonian of electrodynamics in the radiation gauge $H_{\rm rad}=\frac{1}{2\epsilon_{0}}\int d_{3}\mathbf{r}(\boldsymbol{\Pi}^{\dag}\boldsymbol{\Pi}-\epsilon_{0}^{2}c^{2}\mathbf{A}^{\dag}\nabla^2\mathbf{A})$ \cite{COHEN-TANNOUDJI} and the orthogonality of the mode functions $\mathbf{F}_{j}$, we find the cavity Hamiltonian $H_{\rm cav}=\sum_{j}\pi_{j}^2/(2\epsilon_{0})+(\epsilon_{0}\omega_{j}^{2}(z)/2)\xi_{j}^2$, which, of course, takes the form of the Hamiltonian of a set of uncoupled harmonic oscillators. We introduce the normal variables $a_{j}(z)=\sqrt{\epsilon_{0}\omega_{j}(z)/2}\;\xi_{j}+i/\sqrt{2\epsilon_{0}\omega_{j}(z)}\;\pi_{j}$ and their complex conjugates $a_{j}^{\ast}(z)$, so that $H_{\rm cav}=\sum_{j}\omega_{j}(z)|a_{j}(z)|^2$. We restrict ourselves to two cavity modes and assume that the others are sufficiently off-resonant so that they can be neglected. As discussed in the main text, the modes are separated by a free spectral range of the cavity, which we assume is much smaller than both $\omega_{1}^{(0)}$ and $\omega_{2}^{(0)}$, but much larger than any other relevant time scale. It follows that the frequency difference $|\omega_{1}^{(0)}-\omega_{2}^{(0)}|$ can be neglected with respect to the absolute frequencies $\omega_{1}^{(0)}$ and $\omega_{2}^{(0)}$ but that the modes can be addressed independently. The laser drives are modeled by $H_{\rm drive}=\sum_{j=1,2}\Omega_{j}(a_{j}(z)+a_{j}^{\ast}(z))(e^{i\omega_{j}^{(d)}t}+e^{-i\omega_{j}^{(d)}t})$. We move to a frame rotating with the drive frequencies $a\rightarrow a_{j} e^{i\omega_{j}^{(d)}t}$, neglect terms rotating with $\pm 2\omega_{j}^{(d)}$ and assume that $g<<\omega$ so that the $z$-dependence of $a$ can be neglected, while we keep that of the detunings $\Delta_{1,2}(z)=\omega^{(d)}_{1,2}-\omega_{1,2}^{(0)}(z)$. Thus, we obtain the full Hamiltonian in Eq. (\ref{fullham}). So far we have assumed that $z$ is fixed but this analysis applies to $z(t)$, provided that the time dependence is adiabatic with respect to the optical time scales.

\bibliographystyle{prsty}

\end{document}